\documentclass[sigconf]{acmart}
\usepackage[per-mode=symbol,per-symbol=p]{siunitx}
\usepackage[utf8]{inputenc}
\usepackage{adjustbox}
\usepackage{multirow}
\usepackage[ruled,vlined]{algorithm2e}
\usepackage{algorithmic}
\usepackage{graphicx}
\usepackage{textcomp}
\usepackage{xcolor}
\usepackage{balance}
\usepackage{enumitem}
\usepackage{setspace}
\usepackage{wrapfig}
\usepackage{listings}
\usepackage{color}
\usepackage{caption}
\usepackage{subcaption}
\usepackage{makecell}

\usepackage{amsmath}
\DeclareMathOperator*{\argmax}{arg\,max}
\DeclareMathOperator*{\argmin}{arg\,min}
\usepackage{mathtools}

\newcommand{\etal}{\emph{et~al.}\xspace}
\newcommand{\ie}{\emph{i.e.}, }
\newcommand{\eg}{\emph{e.g.}, }

\newcommand{\HAS}{\emph{HTTP Adaptive Streaming }}

\newcommand{\HEVC}{\emph{High Efficiency Video Coding }}

\newcommand{\VVC}{\emph{Versatile Video Coding }}

\newcommand{\HLS}{\emph{HTTP Live Streaming }}

\newcommand{\opte}{\texttt{OPTE}\xspace}
\newcommand{\scheme}{\texttt{LADRE}\xspace}

\newcommand{\EY}{$E_{\text{Y}}$}
\newcommand{\EU}{$E_{\text{U}}$}
\newcommand{\EV}{$E_{\text{V}}$}
\newcommand{\LY}{$L_{\text{Y}}$}
\newcommand{\LU}{$L_{\text{U}}$}
\newcommand{\LV}{$L_{\text{V}}$}
\newcommand{\h}{$h$}

\newcommand{\BDRP}{$BDR_{\text{P}}$}
\newcommand{\BDRV}{$BDR_{\text{V}}$}

\copyrightyear{2024}
\acmYear{2024}
\setcopyright{rightsretained}
\acmConference[MHV '24]{Mile-High Video Conference}{February 11--14, 2024}{Denver, CO, USA}
\acmBooktitle{Mile-High Video Conference (MHV '24), February 11--14, 2024, Denver, CO, USA}
\acmDOI{10.1145/3638036.3640801}
\acmISBN{979-8-4007-0493-2/24/02}

\begin{document}

\title{Energy-efficient Adaptive Video Streaming with Latency-Aware Dynamic Resolution Encoding}

\author{Vignesh V Menon}
\email{vignesh.menon@hhi.fraunhofer.de}
\orcid{0000-0003-1454-6146}
\affiliation{
  \institution{\small{Video Communication and Applications Dept.}}
  \institution{Fraunhofer HHI}
  \city{Berlin}
  \country{Germany}
}

\author{Amritha Premkumar}
\email{amritha.premkumar@ieee.org}
\orcid{0009-0006-1480-4984}
\affiliation{
  \institution{\small{Department of Computer Science}}
  \institution{Rheinland-Pfälzische Technische
  Universität}
  \city{Kaiserslautern}
  \country{Germany}
}

\author{Prajit T Rajendran}
\email{prajit.thazhurazhikath@cea.fr}
\orcid{0000-0002-8283-9891}
\affiliation{
  \institution{\small{CEA, List, F-91120 Palaiseau}}
  \institution{Université Paris-Saclay}
  \city{Paris}
  \country{France}
}

\author{Adam Wieckowski}
\email{adam.wieckowski@hhi.fraunhofer.de}
\orcid{0000-0003-0490-5803}
\affiliation{
  \institution{\small{Video Communication and Applications Dept.}}
  \institution{Fraunhofer HHI}
  \city{Berlin}
  \country{Germany}
}

\author{Benjamin Bross}
\email{benjamin.bross@hhi.fraunhofer.de}
\orcid{0000-0002-1608-3774}
\affiliation{
  \institution{\small{Video Communication and Applications Dept.}}
  \institution{Fraunhofer HHI}
  \city{Berlin}
  \country{Germany}
}

\author{Christian Timmerer}
\email{christian.timmerer@aau.at}
\orcid{0000-0002-0031-5243}
\affiliation{
  \institution{\small{Institute of Information Technology}}
  \institution{Alpen-Adria-Universität}
  \city{Klagenfurt}
  \country{Austria}
}

\author{Detlev Marpe}
\email{detlev.marpe@hhi.fraunhofer.de}
\orcid{0000-0002-5391-3247}
\affiliation{
  \institution{\small{Video Communication and Applications Dept.}}
  \institution{Fraunhofer HHI}
  \city{Berlin}
  \country{Germany}
}

\renewcommand{\shortauthors}{Vignesh V Menon~\etal}

\begin{abstract}
Traditional per-title encoding schemes aim to optimize encoding resolutions to deliver the highest perceptual quality for each representation. However, keeping the encoding time within an acceptable threshold for a smooth user experience is important to reduce the carbon footprint and energy consumption on encoding servers in video streaming applications. Toward this realization, we introduce an encoding \underline{l}atency-\underline{a}ware \underline{d}ynamic \underline{r}esolution \underline{e}ncoding scheme (\scheme) for adaptive video streaming applications. \scheme determines the encoding resolution for each target bitrate by utilizing a random forest-based prediction model for every video segment based on spatiotemporal features and the acceptable target latency. Experimental results show that \scheme achieves an overall average quality improvement of \SI{0.58}{\decibel} PSNR and \SI{0.43}{\decibel} XPSNR while maintaining the same bitrate, compared to the HTTP Live Streaming (HLS) bitrate ladder encoding of \SI{200}{\second} segments using the VVenC encoder, when the encoding latency for each representation is set to remain below the \SI{200}{\second} threshold. This is accompanied by a \SI{84.17}{\percent} reduction in overall encoding energy consumption.
\end{abstract}




\maketitle

\section{Introduction}
\HAS (HAS) has become the \textit{de-facto} standard in delivering video content for various clients regarding internet speeds and device types. The fundamental idea behind HAS is to divide the video content into segments and encode each segment at various bitrates and resolutions, called \textit{representations}, stored in plain HTTP servers. These representations continuously adapt the video delivery to the network conditions and device capabilities of the client~\cite{DASH_Survey}. Conventionally, a fixed bitrate ladder, \eg \HLS (HLS) bitrate ladder~\cite{HLS_ladder_ref}, is used in video streaming applications.

\begin{figure*}[t]
\centering
\begin{subfigure}{0.27\textwidth}
    \centering
    \includegraphics[clip,width=\textwidth]{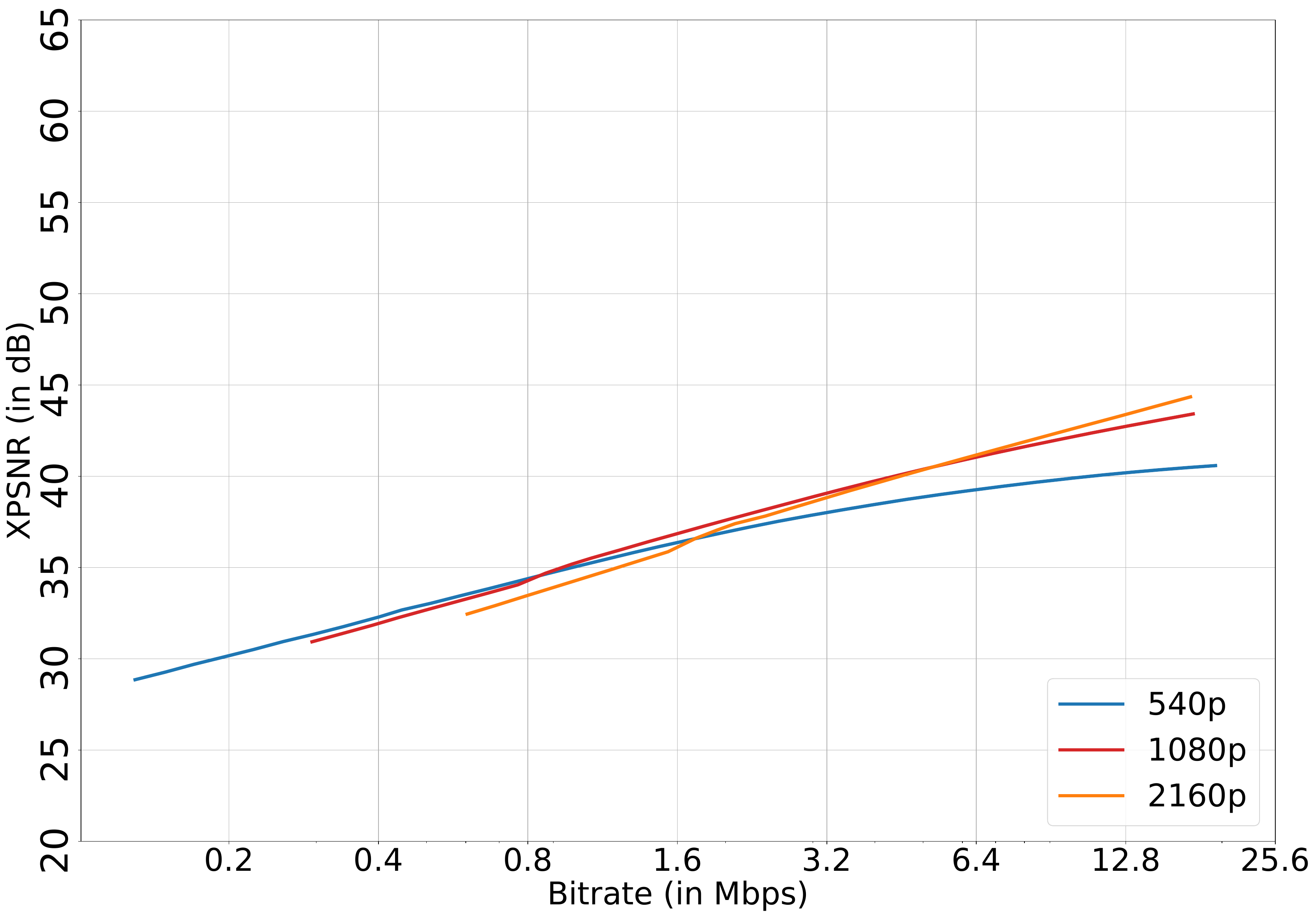}
    \includegraphics[clip,width=\textwidth]{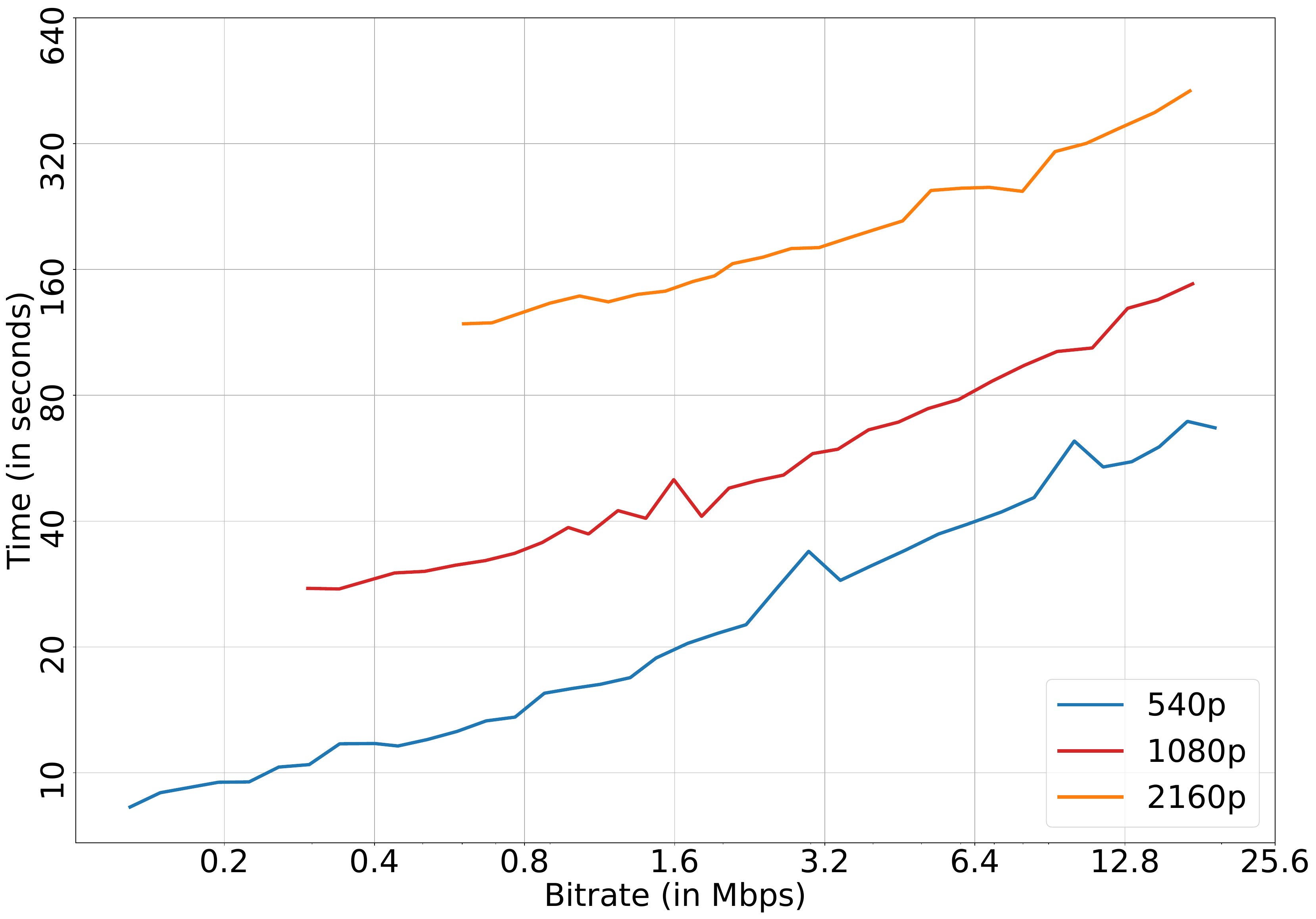}
    \caption{0001 ($E_{\text{Y}}$=20.20, $h$=15.74, $L_{\text{Y}}$=115.96)}    
    \label{fig:bunny_intro}
\end{subfigure}
\hfill
\begin{subfigure}{0.27\textwidth}
    \centering
    \includegraphics[clip,width=\textwidth]{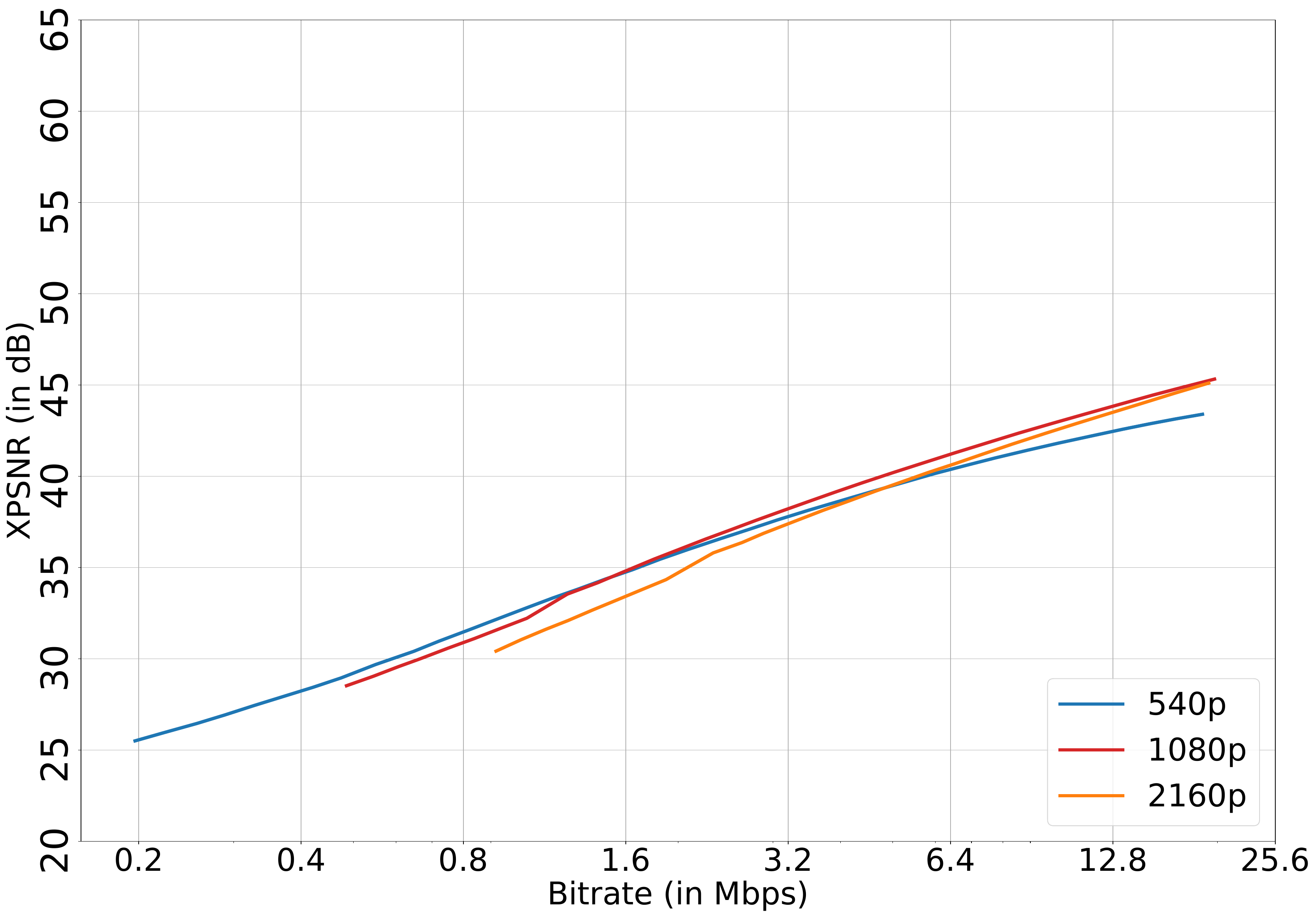}
    \includegraphics[clip,width=\textwidth]{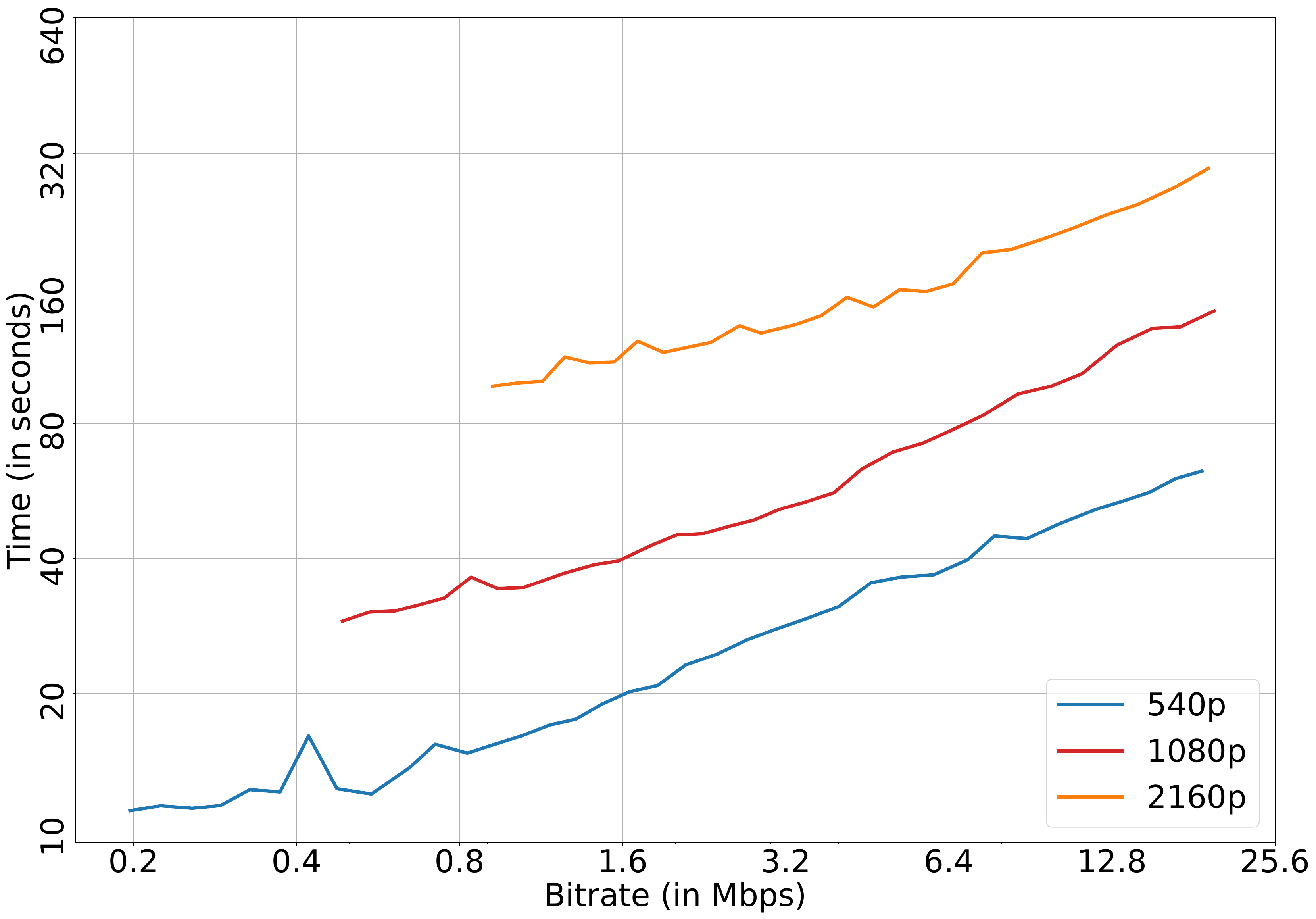}
    \caption{0004 ($E_{\text{Y}}$=55.69, $h$=68.50, $L_{\text{Y}}$=89.34)}    
    \label{fig:honey_intro}
\end{subfigure}
\hfill
\begin{subfigure}{0.27\textwidth}
    \centering
    \includegraphics[clip,width=\textwidth]{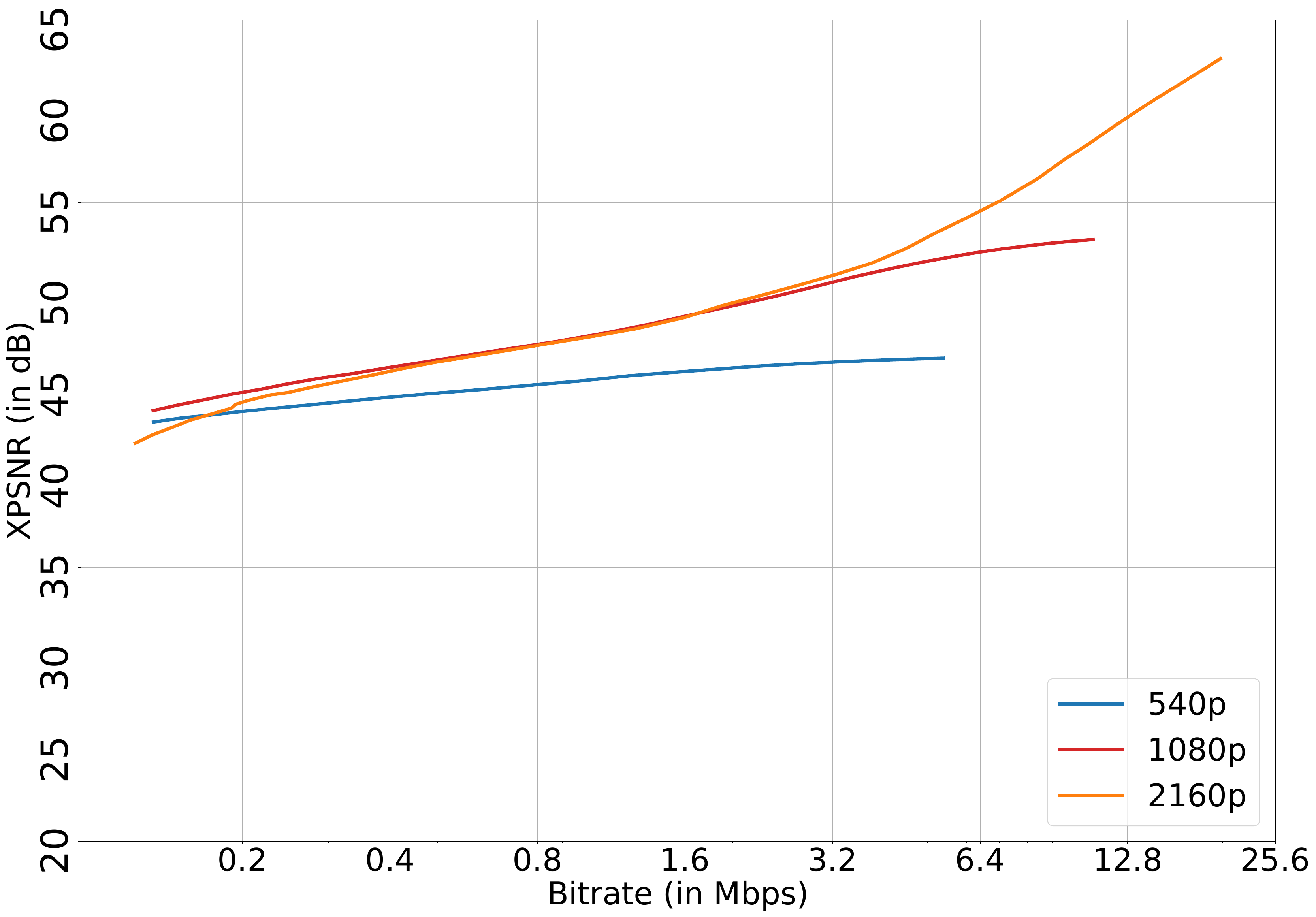}
    \includegraphics[clip,width=\textwidth]{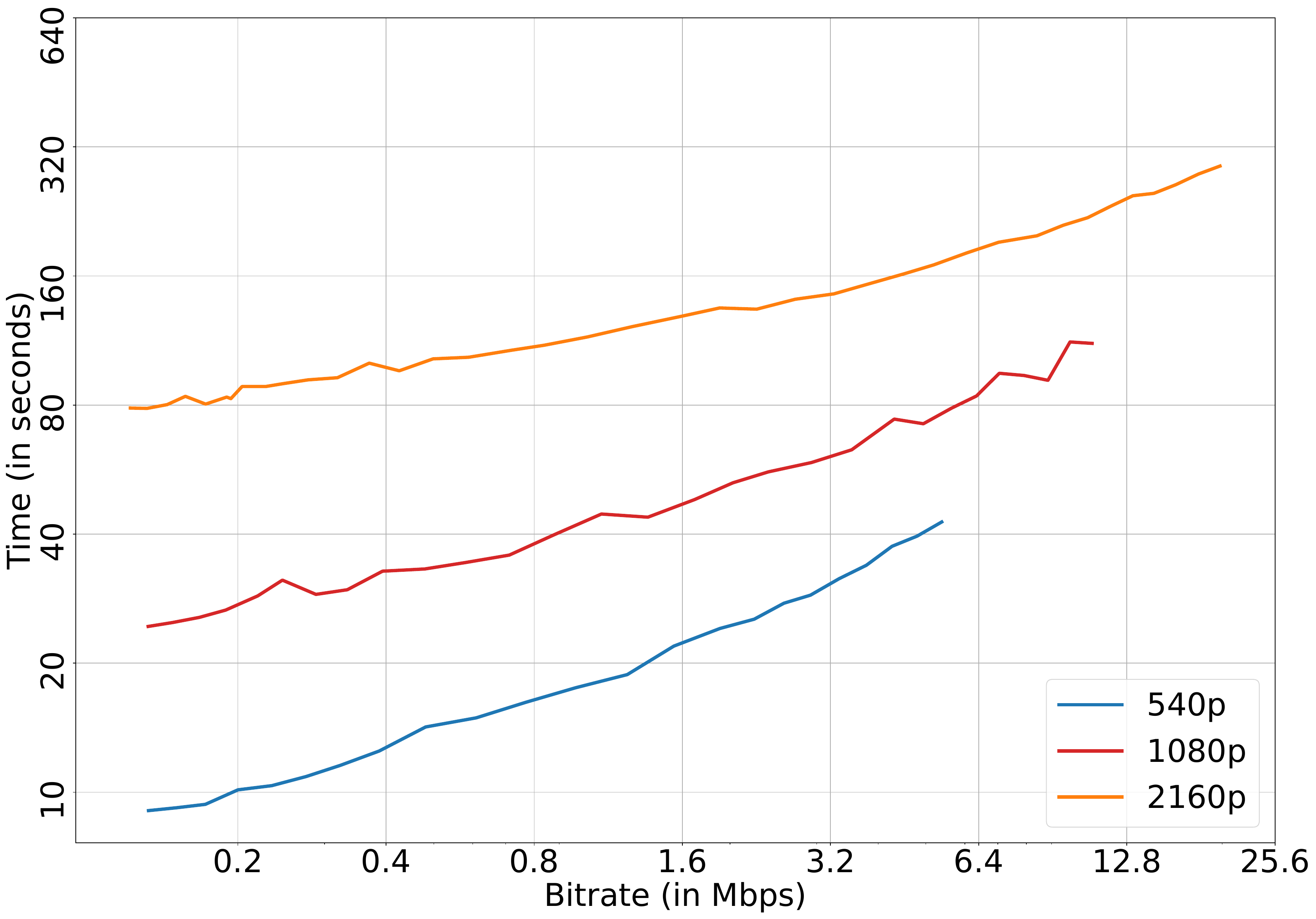}
    \caption{0010 ($E_{\text{Y}}$=1.00, $h$=0.62, $L_{\text{Y}}$=108.98)}    
    \label{fig:0010_intro}
\end{subfigure}
\caption{Rate-distortion (RD) and rate-encoding time curves of representative sequences (segments) of Inter-4K dataset~\cite{inter4k_ref} encoded at 540p, 1080p and 2160p resolutions using VVenC at \textit{faster} preset. Here, XPSNR is used as the quality metric.}
\label{fig:intro_convexhull}
\end{figure*}

\textit{\textbf{Dynamic resolution per-title encoding: }}
In contemporary streaming applications, there is growing interest in per-title encoding methods to enhance the perceived quality of delivered content~\cite{netflix_paper}. This innovative approach dynamically adjusts the encoding resolution~\cite{netflix_paper, gnostic, res_pred_ref1,jtps_ref} and the framerate~\cite{cvfr_ref}, preset~\cite{nasiri_multi_preset,jale_ref}, and other parameters in response to content complexity and viewer preferences to maximize visual fidelity. Dynamic resolution encoding stands out as the most extensively studied per-title encoding scheme in adaptive streaming applications, focusing on adjusting encoding resolutions dynamically to optimize video quality. Adapting the encoding resolution for each segment ensures that the video maintains high perceptual quality in visually intricate segments while efficiently lowering resolution in less complex scenes. As illustrated by the rate-distortion plots of representative segments in Figure~\ref{fig:intro_convexhull} (based on the Inter-4K dataset~\cite{inter4k_ref}), the optimized resolution, which yields the highest perceptual quality (in terms of XPSNR~\cite{xpsnr_ref,xpsnr_itu_ref}), depends on the content complexity. For sequence \textit{0001}, 540p yields the highest XPSNR until 0.8\,Mbps, while 1080p yields the highest XPSNR within a bitrate range of 0.8\,Mbps and 5.0\,Mbps. 2160p yields the highest XPSNR for bitrates over 5.0\,Mbps. Meanwhile, for sequence \textit{0010}, 1080p yields the highest XPSNR until 1.6\,Mbps, while 2160p yields the highest XPSNR for bitrates over 1.6\,Mbps. The streaming system can allocate resources effectively by tailoring the resolution per segment, prioritizing high-quality representations where it matters most. Ultimately, dynamic resolution per-title encoding strives to balance perceptual quality and bandwidth efficiency, offering viewers an immersive and engaging streaming experience~\cite{netflix_paper}. Notably, a convex hull will differ from one codec to the other since the more efficient the codec, the less bitrate it needs to encode at a specific resolution with acceptable quality. For example, \HEVC~(HEVC)~\cite{HEVC} and \VVC~(VVC)~\cite{vvc_ref} allow for better use of the higher resolution than the other codecs by providing a wider range of bitrates to stream that resolution~\cite{vvc_convexhull_ref}.

\textit{\textbf{Encoding delay tolerance in video streaming: }}
In live sports or events streaming, where real-time viewing is crucial, encoding delays are typically expected to be very low, usually in a few seconds or less. This minimal delay ensures that viewers receive the action as close to real-time as possible~\cite{pradeep_ref}. For interactive content like online gaming or live seminars, extremely low encoding delays are crucial in maintaining real-time interactivity. Delays of a few seconds or less are generally acceptable to keep interactions seamless. News broadcasts require relatively low encoding delays to deliver the latest updates. Delays of ten seconds or less are commonly permitted to ensure viewers receive updates as events unfold. Reducing encoding time is also critical in video-on-demand (VoD) streaming applications since it contributes to environmental sustainability. Encoding processes in data centers require substantial computational resources and energy consumption. The streaming industry can reduce its carbon footprint and energy consumption by minimizing encoding time. This is particularly important as environmental consciousness grows and companies seek to adopt greener practices. Lower encoding latency aligns with eco-friendly streaming practices by optimizing resource utilization and minimizing unnecessary energy expenditure.

In this light, encoding time-aware per-title encoding is essential for adaptive streaming applications. Firstly, it enables the adaptation of encoding parameters based on the time constraints of the encoding process. By considering encoding time as a crucial factor, the encoding settings can be optimized dynamically to ensure that content is encoded within specific time limits, meeting the demands of real-time streaming scenarios. Secondly, it enhances resource allocation efficiency by tailoring the encoding parameters to match the available computational resources, thereby avoiding overloading or underutilizing encoding servers. Additionally, it contributes to an overall improvement in streaming efficiency, allowing for smoother content delivery by minimizing encoding delays and ensuring that the streaming service can efficiently handle varying workloads and complexities of encoding tasks. Encoding time-aware per-title encoding is pivotal in optimizing streaming quality while efficiently utilizing computational resources in online adaptive streaming environments.

As shown in the rate-encoding time plots of representative segments in Figure~\ref{fig:intro_convexhull}, the encoding time depends on the encoding resolution chosen for the video content. This relationship is intrinsic to the encoding process, where the number of pixels in each frame significantly impacts the computational workload. Higher encoding resolutions, such as 2160p or 4320p, encompass more pixels, necessitating increased computational resources and time for encoding compared to lower resolutions like 720p or 1080p. The dependency on encoding resolution becomes especially relevant in adaptive streaming, where content is encoded at multiple resolutions to accommodate varying user devices and network conditions.  However, state-of-the-art per-title encoding methods do not consider the encoding latency constraints while optimizing the encoding resolution~\cite{netflix_paper,faust_ref,gnostic,jtps_ref}. 

\textit{\textbf{Contributions:}} 
Our main contributions are as follows:
\begin{enumerate}
\item An online encoding resolution selection algorithm to maximize the perceived quality of video segments based on their spatiotemporal complexity, target bitrate, and the encoding time constraint. 
\item Implementation of the proposed per-title encoding scheme using an open-source VVC-based toolchain.
\item Comprehensive analysis of the proposed resolution selection algorithm for various encoding time thresholds regarding compression efficiency and encoding latency. Since no open-source VVC encoder implementations are available for live encoding, we limit our evaluation to the VoD streaming scenario.
\end{enumerate}

\textit{\textbf{Outline: }}
The remainder of this paper is organized as follows. Section~\ref{sec:dyn_res} discusses related work on dynamic resolution encoding in the context of adaptive video streaming. The proposed \scheme scheme is introduced and described in detail in Section~\ref{sec:prop_framework}. Section~\ref{sec:exp_design} explains the experimental design, while Section~\ref{sec:results} presents the experimental results. Finally, Section~\ref{sec:conclusion} concludes the paper.

\section{Related work}
\label{sec:dyn_res}
Most state-of-the-art dynamic resolution per-title encoding methods are based on choosing a particular resolution that provides better visual quality for a given bitrate range. Table~\ref{tab:num_pre_enc} shows the target scenario, the bitrate estimation method, the number of pre-encodings needed to determine the convex hull, and the encoding type of the state-of-the-art methods. 
\begin{table}[t]
\caption{Comparison of the state-of-the-art dynamic resolution per-title encoding methods with \scheme.}
\centering
\resizebox{0.85\linewidth}{!}{
\begin{tabular}{l|c|c|c}
\specialrule{.12em}{.05em}{.05em}
\specialrule{.12em}{.05em}{.05em}
Method & \makecell{Number of\\ pre-encodings} & \makecell{Encoding\\ type} & \makecell{Encoding latency\\ awareness} \\
\specialrule{.12em}{.05em}{.05em}
\specialrule{.12em}{.05em}{.05em}
Bruteforce~\cite{netflix_paper, chen_pte_ref} & $\Tilde{r} \times \Tilde{c}$ & cVBR & No \\
Katsenou~\etal~\cite{gnostic}  & $(\Tilde{r} - 1) \times 2$ & CQP & No \\
\texttt{FAUST}\cite{faust_ref} & 1 & CBR & No \\
Bhat~\etal~\cite{res_pred_ref1} & 1 & CBR & No \\
\texttt{OPTE}\cite{jtps_ref} &  0 & cVBR & No\\
\specialrule{.12em}{.05em}{.05em}
\scheme & \textbf{0} & \textbf{cVBR} & \textbf{Yes} \\
\specialrule{.12em}{.05em}{.05em}
\specialrule{.12em}{.05em}{.05em}
\end{tabular}
}
\vspace{-0.7em}
\label{tab:num_pre_enc}
\end{table}
Katsenou~\etal~\cite{gnostic} uses machine learning to identify the most effective bitrate range for each resolution. The method extracts spatiotemporal features and statistics from sequences at their original resolution. Then, it employs machine learning methods to predict the quantization parameters (QPs) at which the rate-distortion curves across the different resolutions intersect. $(\Tilde{r}-1)\times 2$ encodes must be performed to determine the bitrates at which resolutions should be switched.  This content-gnostic approach has been claimed to reduce the number of encodings required compared to other methods (by 81\% - 94\%) compared to the bruteforce encoding approach. It uses constant quantization parameter (CQP) encodes, which are not used in real-time streaming applications. 
Another method proposed by Bhat~\etal~\cite{res_pred_ref1} uses machine learning to predict the resolution without requiring multiple encodings. Features from the low-resolution encoding of the first few frames are input to a random forest model to predict better-performing resolution for a decision period. Similarly, Zabrovskiy~\etal~\cite{faust_ref} used an artificial neural network to predict an optimized bitrate ladder for each scene, optimized based on the YPSNR quality metric. These methods produce \textit{latency} significantly higher than the accepted latency in live streaming.
Our previous work \texttt{OPTE}~\cite{jtps_ref} uses random forest models to predict optimized resolution, yielding the highest perceptual quality using spatiotemporal features extracted for each segment. However, \opte does not consider encoding latency constraint during the optimized resolution prediction. To summarize, current related work lacks considering encoding latency constraints while selecting the optimized encoding resolution, and most state-of-the-art methods need pre-encodings which yield significant latency and energy consumption.

\section{Latency-aware dynamic resolution encoding}
\label{sec:prop_framework}
\begin{figure*}[t]
\centering
\includegraphics[width=\linewidth]{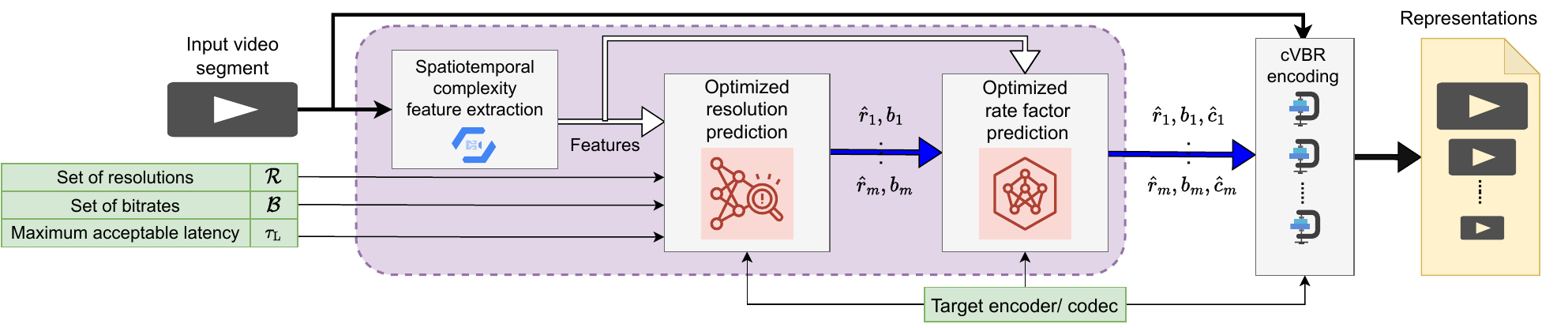}
\caption{Encoding using \scheme envisioned in this paper for online streaming applications.}
\label{fig:contribution}
\end{figure*}

Striking the right balance between offering high-quality, high-resolution streams and minimizing encoding time and energy consumption is crucial for adaptive streaming platforms to ensure responsive and uninterrupted playback experiences across various end-user devices and network environments. In line with this perspective, this paper proposes an encoding latency-aware dynamic encoding resolution encoding scheme (\scheme) to maximize the perceived quality of video segments (in terms of XPSNR) based on the video content complexity, target bitrate, and the maximum encoding time constraint. As shown in Figure~\ref{fig:contribution}, \scheme is classified into four steps and described in the following subsections:
\begin{enumerate}
    \item spatiotemporal complexity feature extraction,
    \item optimized resolution prediction,
    \item optimized rate factor prediction, and
    \item constrained variable bitrate (cVBR) encoding using the selected bitrate-resolution-rate factor combinations.
\end{enumerate}
Table~\ref{tab:notation} summarizes the notations used in \scheme.

\begin{table}[t]
\centering
\caption{Notations used in \scheme.}
\label{tab:notation}
\resizebox{\linewidth}{!}{
\begin{tabular}{c|l}
\specialrule{.12em}{.05em}{.05em}
\specialrule{.12em}{.05em}{.05em}
\emph{Notation} & \emph{Description}\\
\specialrule{.12em}{.05em}{.05em}
\specialrule{.12em}{.05em}{.05em}
\multicolumn{2}{c}{\emph{Video complexity features}}\\\hline
\EY & Average luma texture energy of segment\\
\h & Average gradient of the luma texture energy of segment \\
\LY & Average luminescence of segment \\
\EU, \EV & Average chroma texture energy of segment (U and V channels) \\
\LU, \LV & Average chrominescence of segment (U and V channels) \\
\hline
\multicolumn{2}{c}{\emph{Input parameters}}\\\hline
$\mathcal{R}$ & Set of supported resolutions\\
$\mathcal{B}$ & Set of supported bitrates\\
$\tau_{\text{L}}$ & Maximum acceptable encoding latency\\
\hline
$\hat{r}_{t},\hat{b}_{t},\hat{c}_{t}$ & predicted resolution, bitrate and rate factor of the $t$\textsuperscript{th} representation\\
\specialrule{.12em}{.05em}{.05em}
\specialrule{.12em}{.05em}{.05em}
\end{tabular}}
\end{table}

\subsection{Spatiotemporal complexity feature extraction}
\label{sec:features}
This process involves analyzing the video content in both spatial and temporal dimensions, capturing essential information about object movements, scene changes, and visual details. Predictive models can comprehensively understand the content complexity and characteristics by extracting relevant spatiotemporal features, such as motion vectors, texture patterns, and frame-to-frame differences~\cite{vqtif_ref}. These features serve as valuable inputs for algorithms that dynamically adjust encoding parameters like bitrate and resolution, ensuring that the live stream maintains optimal quality while adapting to real-time changing network conditions. Intuitively, a higher resolution may be warranted to effectively capture fine details for content with high spatial and relatively low temporal complexity. Conversely, content with high temporal complexity may benefit from a lower resolution, as rapid changes between frames can limit the perceptual benefit of higher spatial detail. This paper uses seven DCT-energy-based features~\cite{dct_ref,vca_ref}: \{\EY, \h, \LY, \EU, \LU, \EV, \LV \} as the content complexity features of video segments.

\subsection{Optimized resolution estimation}
\label{sec:res_pred}
The objective of selecting the optimized resolution based on bitrate and video complexity features is decomposed into two parts:
\begin{enumerate}
    \item designing models to predict the encoding time and the perceptual quality in terms of XPSNR;
    \item developing a function to obtain the optimized resolution based on the predicted encoding times and perceptual quality for each available encoding resolution.
\end{enumerate}

\textit{\textbf{Modeling: }}  
The perceptual quality $v_{(r_{t},b_{t})}$ and encoding time $\tau_{(r_{t},b_{t})}$ of the representation $(r_{t},b_{t})$ rely on the extracted video complexity features (\EY, \h, \LY, \EU, \LU, \EV, \LV), encoding resolution $r_t$, and target bitrate $b_t$~\cite{cvfr_ref,mcbe_ref}:
\begin{align}
\label{eq:v_pred}
    v_{\left(r_{t},b_{t}\right)} &= f_{\text{V}}\left(E_{\text{Y}}, h, L_{\text{Y}}, E_{\text{U}}, L_{\text{U}}, E_{\text{V}}, L_{\text{V}}, r_{t}, b_{t}\right);\\
\label{eq:s_pred}
    \tau_{\left(r_{t},b_{t}\right)} &= f_{\tau}\left(E_{\text{Y}}, h, L_{\text{Y}}, E_{\text{U}}, L_{\text{U}}, E_{\text{V}}, L_{\text{V}}, r_{t}, b_{t}\right).
\end{align}
Spatio-temporal features encapsulate intricate spatial details and temporal dynamics within the video segment and help to assess the video fidelity~\cite{csvt_ref1}. Including resolution, bitrate, framerate, and preset parameters in the models acknowledges the interplay between compression efficiency, temporal smoothness, and spatial clarity in shaping perceived quality. A higher resolution, bitrate, or framerate may improve the quality and increase the file size of the video segment. A slower preset at the same target bitrate can reduce the file size of the video segment. Similarly, a higher resolution, bitrate, framerate, or a slower preset can increase the encoding duration.

\textit{\textbf{Optimization: }} 
\scheme optimizes the perceptual quality of encoded video segments while adhering to real-time processing constraints. It predicts the optimized resolution of the $t$\textsuperscript{th} representation to maximize the compression efficiency while maintaining the encoding time below the threshold $\tau_{\text{L}}$. The optimization function is:
\begin{align}
   \hat{r}_{t} &= \argmax_{r \in \mathcal{R}} \hat{v}_{(r,b_{t})} & c.t. \hspace{1em} \hat{\tau}_{(r,b_{t})} \leq \tau_{\text{L}}.  
\end{align}
where $\hat{v}_{(r,b_{t})}$ and $\hat{\tau}_{(r,b_{t})}$ are the predicted XPSNR and encoding speed of the representation $(r,b_{t})$. This paper considers XPSNR as the perceptual quality measure instead of the popular VMAF metric since the correlation of XPSNR with subjective quality scores is significantly higher than VMAF for VVC-coded bitstreams~\cite{xpsnr_vs_vmaf}.

\subsection{Optimized rate factor estimation}
\label{sec:crf_pred} 
Predicting the rate factor helps ensure consistent video quality throughout the stream. It allows the encoder to allocate bits judiciously, preventing underallocation (resulting in poor quality) or over-allocation (wasting bandwidth) of bits for encoding~\cite{jtps_ref}.

\textit{\textbf{Modeling: }} 
The rate-factor $c_{(r_{t},b_{t})}$  relies on the extracted video complexity features, encoding resolution $r_t$, and target bitrate $b_t$ parameters~\cite{vvenc_qp_pred}:
\begin{align}
\label{eq:c_pred}
    c_{\left(r_{t},b_{t}\right)} &= f_{\text{C}}\left(E_{\text{Y}}, h, L_{\text{Y}}, E_{\text{U}}, L_{\text{U}}, E_{\text{V}}, L_{\text{V}}, r_{t}, b_{t}\right).
\end{align}
Content with intricate details, textures, or sharp edges demands a lower rate factor to represent these features accurately in the encoded video. Similarly, segments with fast motion, frequent scene changes, or dynamic content require a lower rate factor to accurately capture the rapid changes between frames. 

\textit{\textbf{Optimization: }}
The mathematical formulation of the rate factor optimization to yield a bitrate as close to the target bitrate as possible can be expressed as follows: 
\begin{align}
   \hat{c}_{(r,b_{t})} &= \argmin_{c \in [c_{\text{min}},c_{\text{max}}]} \mid \hat{b}_{(r,c)}- b_{t} \mid .  
\end{align}
A loss function measures the deviation between the target and predicted bitrate. The objective is to find the rate factor that minimizes the loss function.

\subsection{Constrained variable bitrate encoding}
\label{sec:cvbr_pred}
Constrained variable bitrate encoding offers dynamic bitrate allocation while imposing upper limits on the bitrate variability. This method enables flexibility in bitrate adjustment, accommodating the complexity of video content by allowing fluctuations within a predefined range. However, unlike traditional VBR methods, constrained VBR sets constraints on the maximum and minimum bitrates, ensuring that the bitrate variations stay within specific bounds. Doing so balances the trade-off between encoding efficiency and consistent quality. 

The encoding uses the predicted bitrate-resolution-rate factor configurations for a given input video segment. $\hat{b}_{t}$ is considered the upper bound of bitrate variability, and $\hat{c}_t$ is the rate factor used for encoding. In VVenC~\cite{vvenc_ref}, the rate factor is specified using the \texttt{qp} option, while the \texttt{maxrate} (easy mode) or \texttt{MaxBitrate} (expert mode) option is used to specify the upper bound of bitrate variability. A reasonable lower bound of the bitrate variability is achieved using the XPSNR-based perceptual QP adaptation in combination with the \texttt{qp} option and does not need to be specified explicitly.

\section{Experimental design}
\label{sec:exp_design}

\subsection{Evaluation setup}
\label{sec:test_methodology}
All experiments are run on a dual-processor server with Intel Xeon Gold 5218R (80 cores, frequency at 2.10 GHz), where each encoding instance uses four CPU threads (\ie $c$ = 4) with multi-threading and x86 SIMD~\cite{x86_simd_ref} optimizations. The Inter-4K dataset~\cite{inter4k_ref} is used to validate the performance of the encoding schemes considered in this paper. Inter-4K is a standard benchmark
for video super-resolution and frame interpolation, containing 1000 UHD videos captured at 60 frames per second (fps). The experimental parameters used to evaluate \scheme are shown in Table~\ref{tab:exp_par}. The sequences are encoded using VVenC v1.10~\cite{vvenc_ref} using preset 0 (\textit{faster}).  The spatiotemporal features are extracted using VCA v2.0~\cite{vca_ref} running as a pre-processor using four CPU threads with multi-threading and x86 SIMD optimizations. \scheme uses random forest regression models~\cite{rf_ref} to predict XPSNR, encoding time, and rate factor trained for each supported resolution in $\mathcal{R}$.

\subsection{Performance metrics}
The resulting overall quality in Peak Signal to Noise Ratio (PSNR) \cite{psnr_ref} and XPSNR~\cite{xpsnr_ref} and the achieved bitrate are compared for every representation of each test sequence. Bjøntegaard Delta values~\cite{DCC_BJDelta} BD-PSNR and BD-XPSNR refer to the average increase in PSNR and XPSNR of the representations compared with the reference encoding scheme with the same bitrate, respectively. A positive BD-PSNR and BD-XPSNR indicate a gain in coding efficiency compared to the reference encoding scheme. 
The relative storage consumption ($\Delta S$) of representations is evaluated compared with the reference encoding scheme. 
The encoding latency of a video segment is calculated from the instant it is fed to the \scheme pipeline until each representation is completely encoded.

\begin{table}[t]
\caption{Experimental parameters used to evaluate \scheme.}
\centering
\resizebox{0.9\columnwidth}{!}{
\begin{tabular}{l||c|c|c|c|c|c}
\specialrule{.12em}{.05em}{.05em}
\specialrule{.12em}{.05em}{.05em}
\emph{Parameter} & \multicolumn{6}{c}{\emph{Values}}\\
\specialrule{.12em}{.05em}{.05em}
\specialrule{.12em}{.05em}{.05em}
$\mathcal{R}$ & \multicolumn{6}{c}{\{ 360, 540, 720, 1080, 1440, 2160 \} } \\
\hline
$\mathcal{B}$ & 0.145 & 0.300 & 0.600 & 0.900 & 1.600 & 2.400 \\
              & 3.400 & 4.500 & 5.800 & 8.100 & 11.600 & 16.800 \\
\hline
 $\tau_{\text{L}}$ & \multicolumn{6}{c}{ \SI{50}{\second}, \SI{100}{\second}, \SI{200}{\second}, \SI{400}{\second}, $\infty$ } \\
\hline
Target encoder & \multicolumn{6}{c}{VVenC (faster)} \\
\hline
CPU threads & \multicolumn{6}{c}{4} \\
\specialrule{.12em}{.05em}{.05em}
\specialrule{.12em}{.05em}{.05em}
\end{tabular}
}
\label{tab:exp_par}
\end{table}

\subsection{Reference schemes}
This paper compares \scheme with the following benchmarks:
\begin{enumerate}
    \item \texttt{Default}:  This scheme employs a fixed bitrate ladder, \ie a fixed set of bitrate-resolution pairs. This paper uses the HLS bitrate ladder specified in the Apple authoring specifications~\cite{HLS_ladder_ref} as the fixed bitrate ladder.
    \item \opte~\cite{jtps_ref} This scheme predicts optimized resolution, which yields the highest XPSNR for a given target bitrate. 
\end{enumerate}
The other methods mentioned in Table~\ref{tab:num_pre_enc} are not considered for evaluation, as they utilize information from pre-encodings to determine convex-hull, which introduces significant latency and energy consumption in the encoding servers.

\section{Evaluation results}
\label{sec:results}

\textit{\textbf{Prediction latency and accuracy:}} We evaluate the pre-processing latency ($\tau_{\text{p}}$) in encoding introduced by the video complexity feature extraction and the model inference to predict the optimized resolution-rate factor configurations. We extract the features at an average rate of 167\,fps over the entire dataset (2160p resolution). This result is critical in future-proofing the system by handling evolving content requirements (\eg 8K resolution or high framerate content). The time to predict the resolution-rate factor for each representation is \SI{5}{\milli\second}. As video complexity feature extraction and the optimized resolution and rate factor prediction can execute concurrently in real applications, the overall latency introduced by \scheme is negligible. The accuracy of the encoding time, rate factor, and XPSNR prediction models are analyzed in terms of mean absolute error (MAE). The average MAE is \SI{6.97}{\second}, 1.05, and \SI{0.48}{\decibel}, respectively.

\begin{figure}[t]
\centering
\begin{subfigure}{0.48\linewidth}
    \centering
    \includegraphics[width=0.98\textwidth]{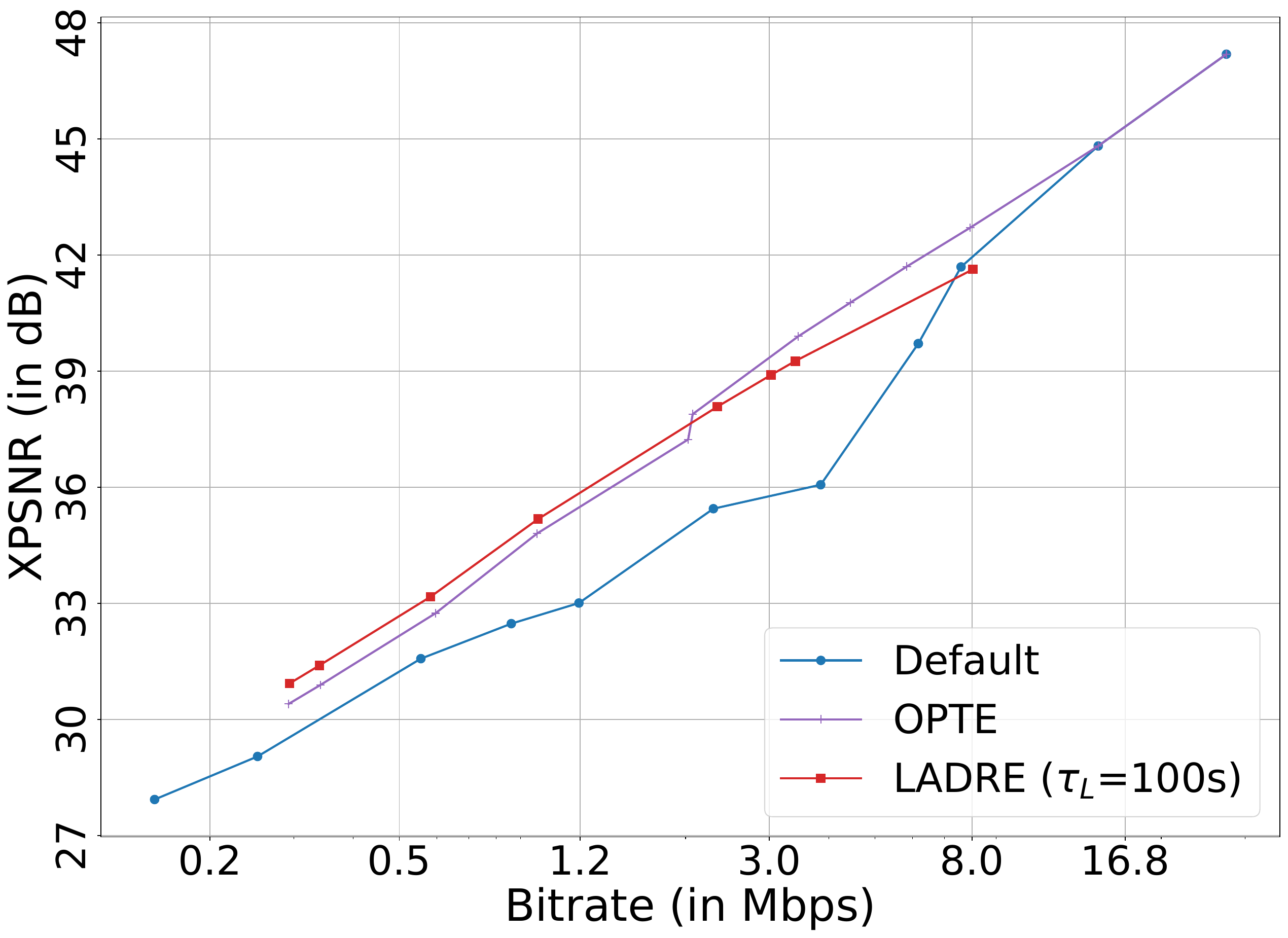}
    \includegraphics[width=0.98\textwidth]{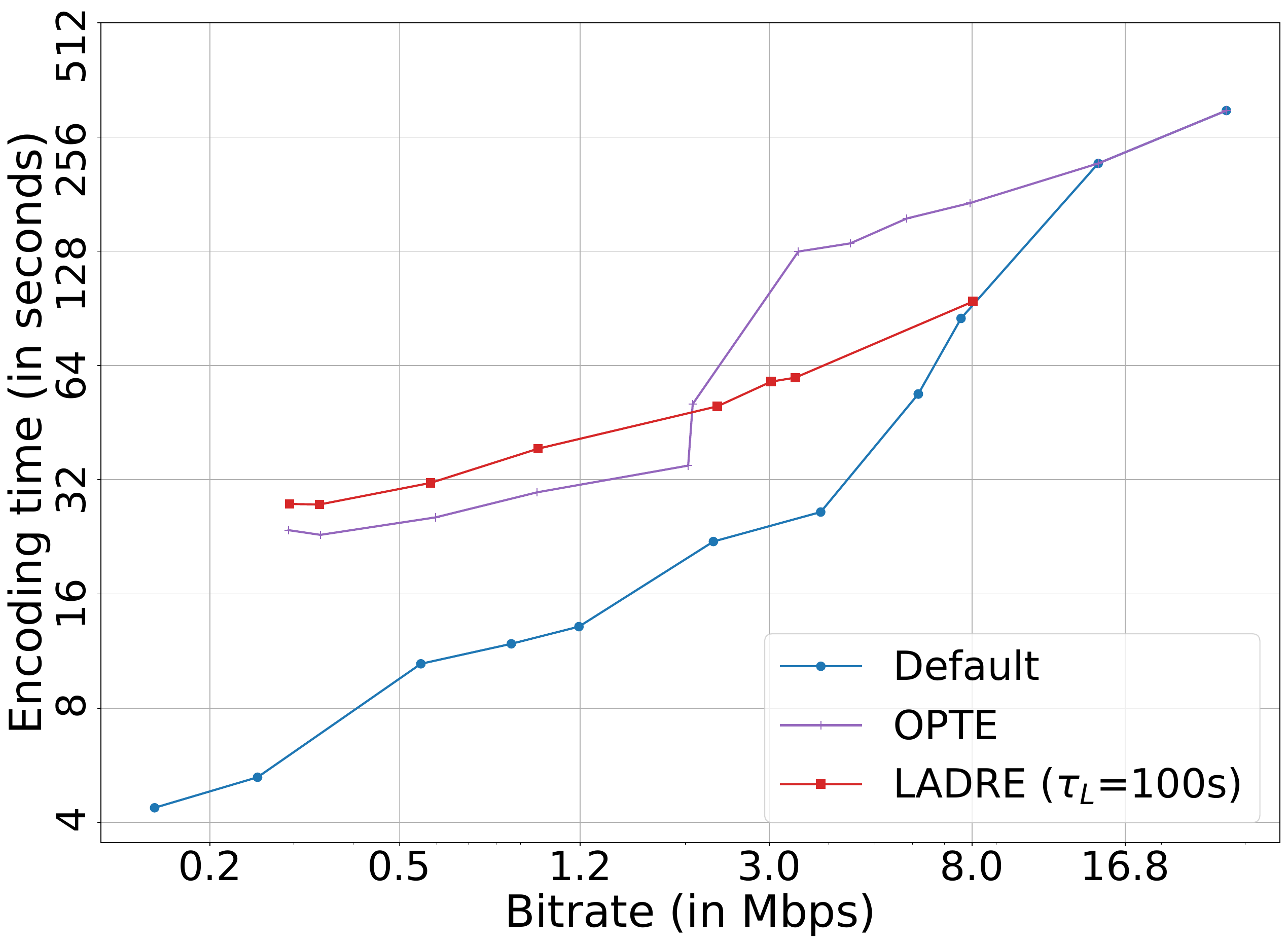}   
    \caption{\textit{0020}}
\end{subfigure}
\begin{subfigure}{0.48\linewidth}
    \centering
    \includegraphics[width=0.98\textwidth]{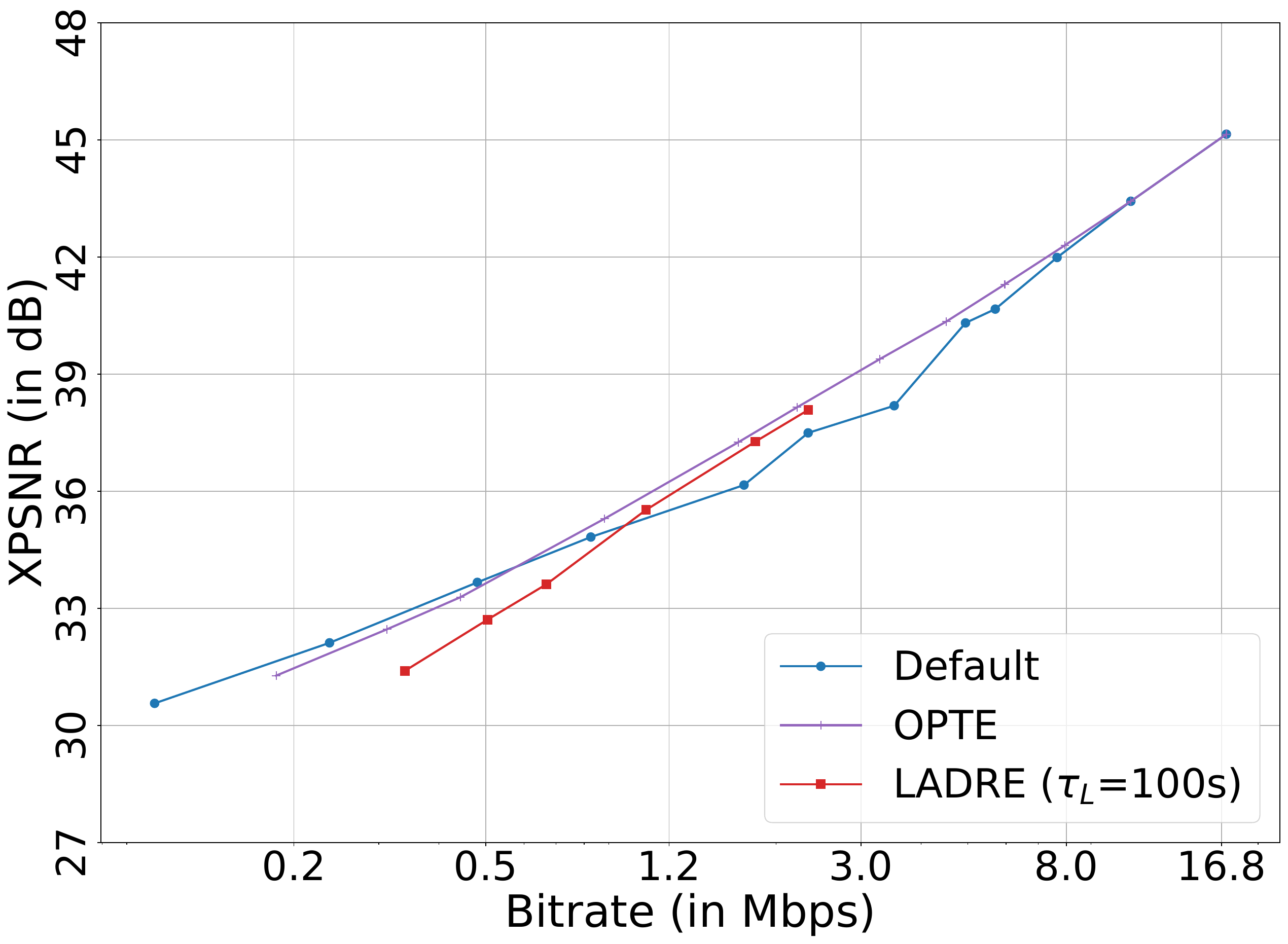}
    \includegraphics[width=0.98\textwidth]{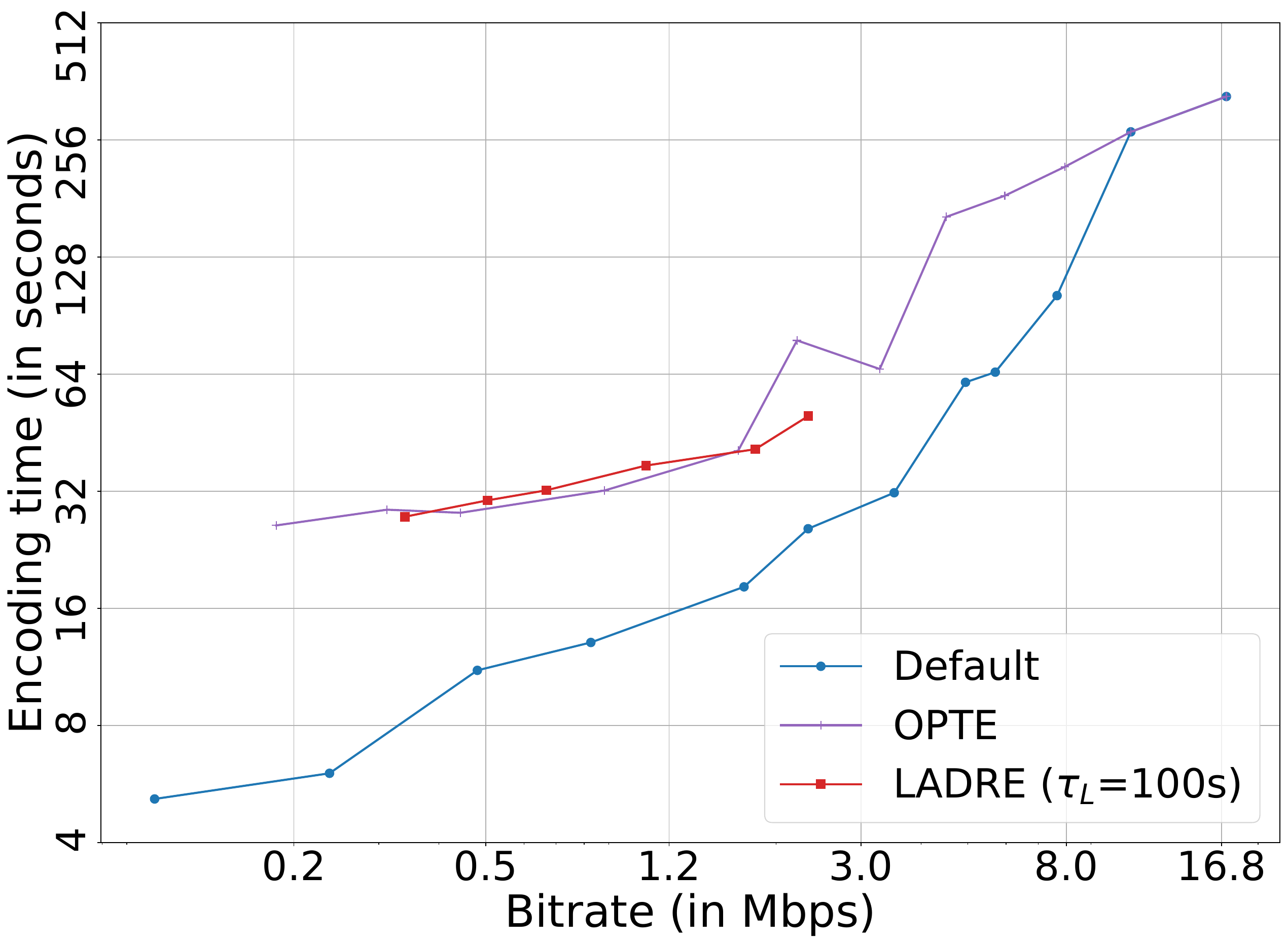}    
    \caption{\textit{0057}}
\end{subfigure}
\begin{subfigure}{0.48\linewidth}
    \centering
    \includegraphics[width=0.98\textwidth]{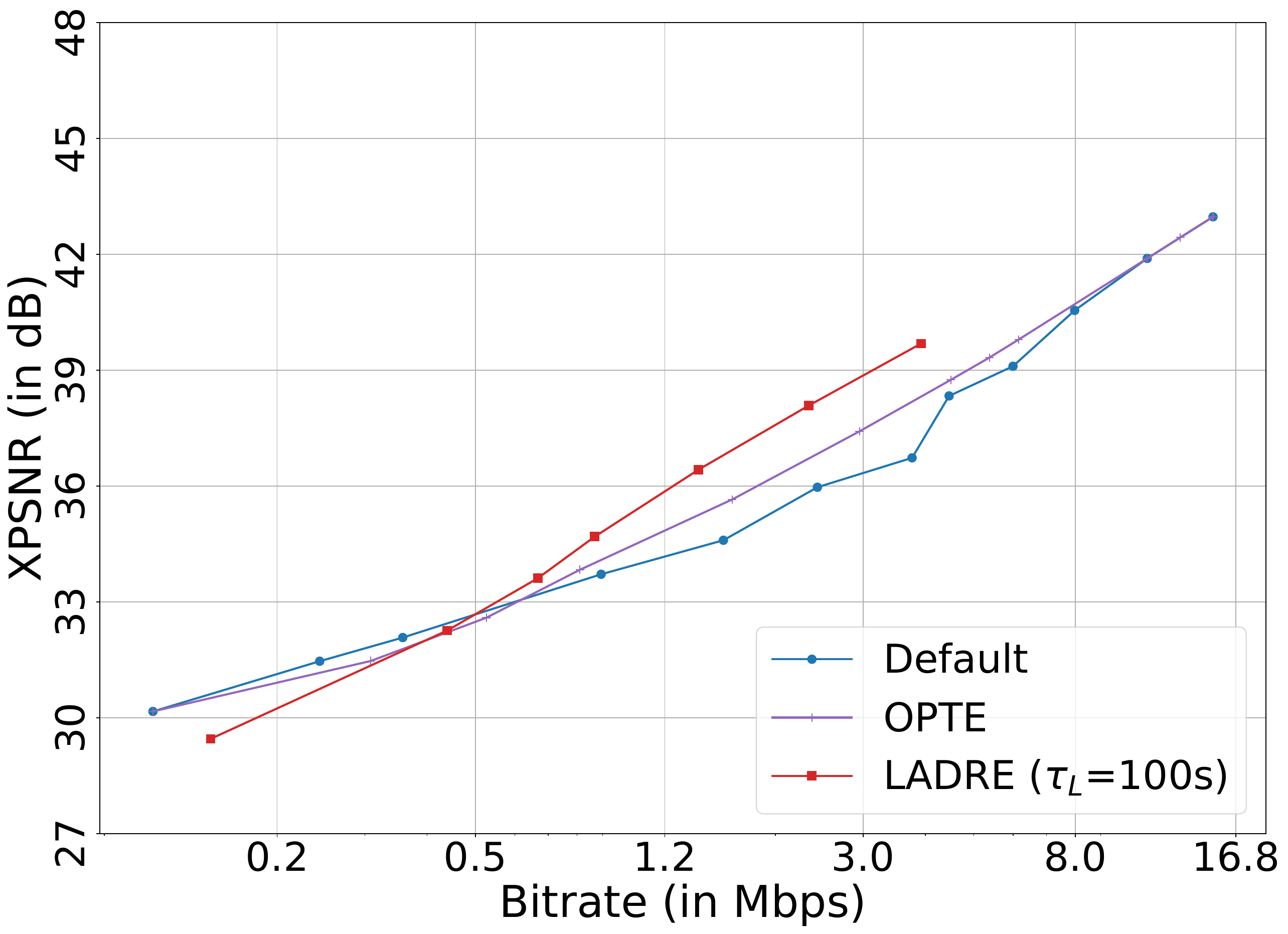}
    \includegraphics[width=0.98\textwidth]{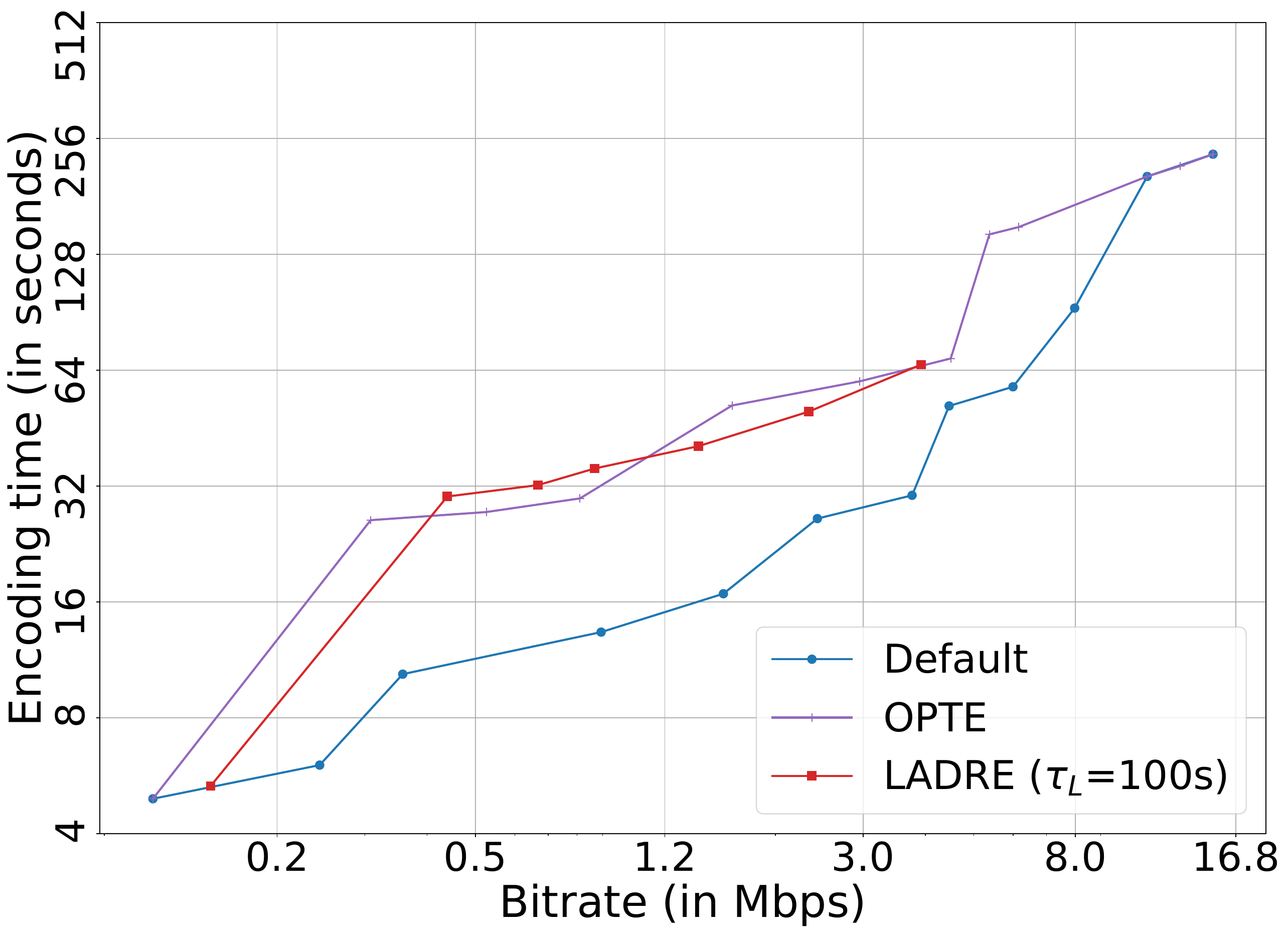}    
    \caption{\textit{0642}}
\end{subfigure}
\begin{subfigure}{0.48\linewidth}
    \centering
   \includegraphics[width=0.98\textwidth]{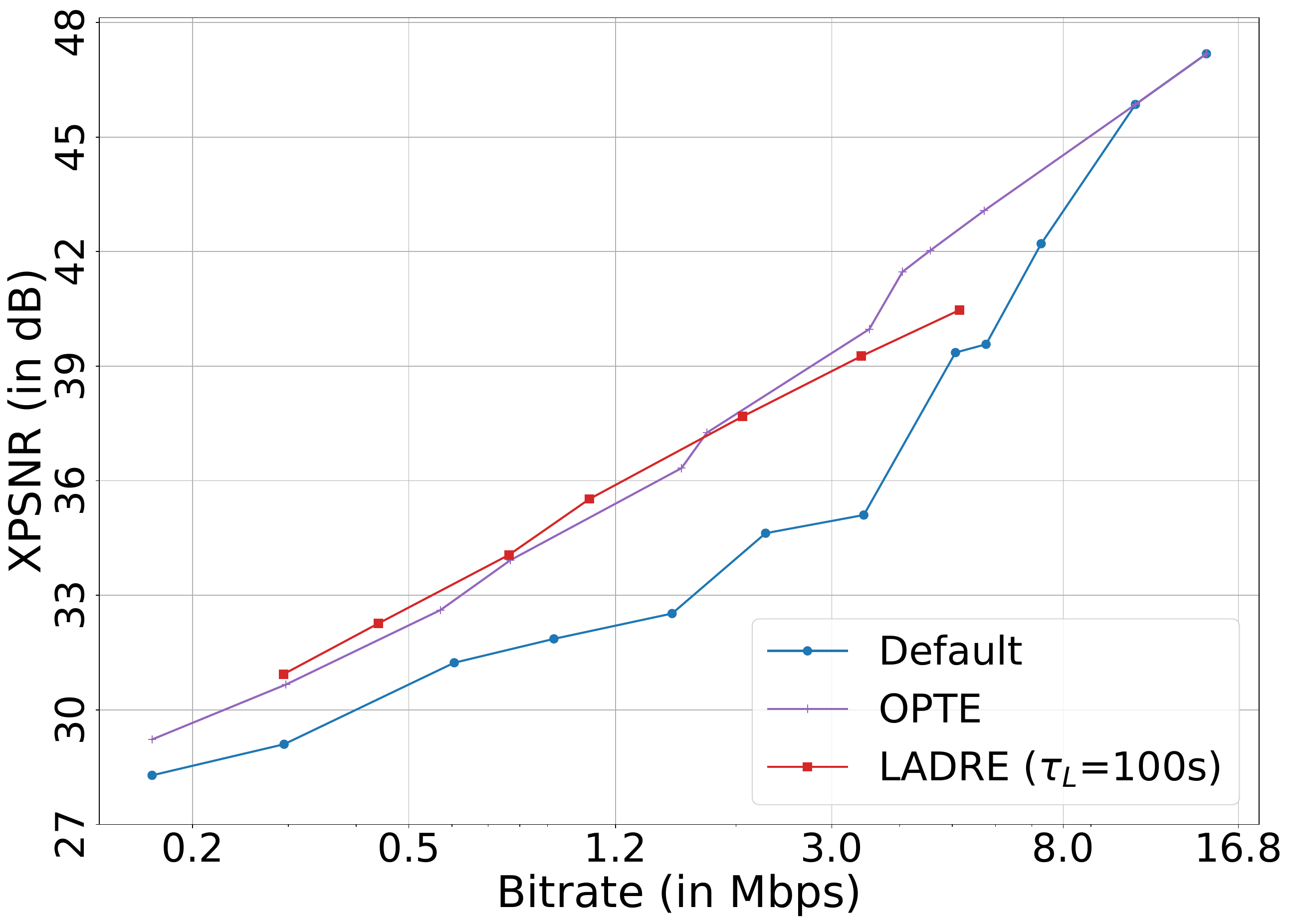}
    \includegraphics[width=0.98\textwidth]{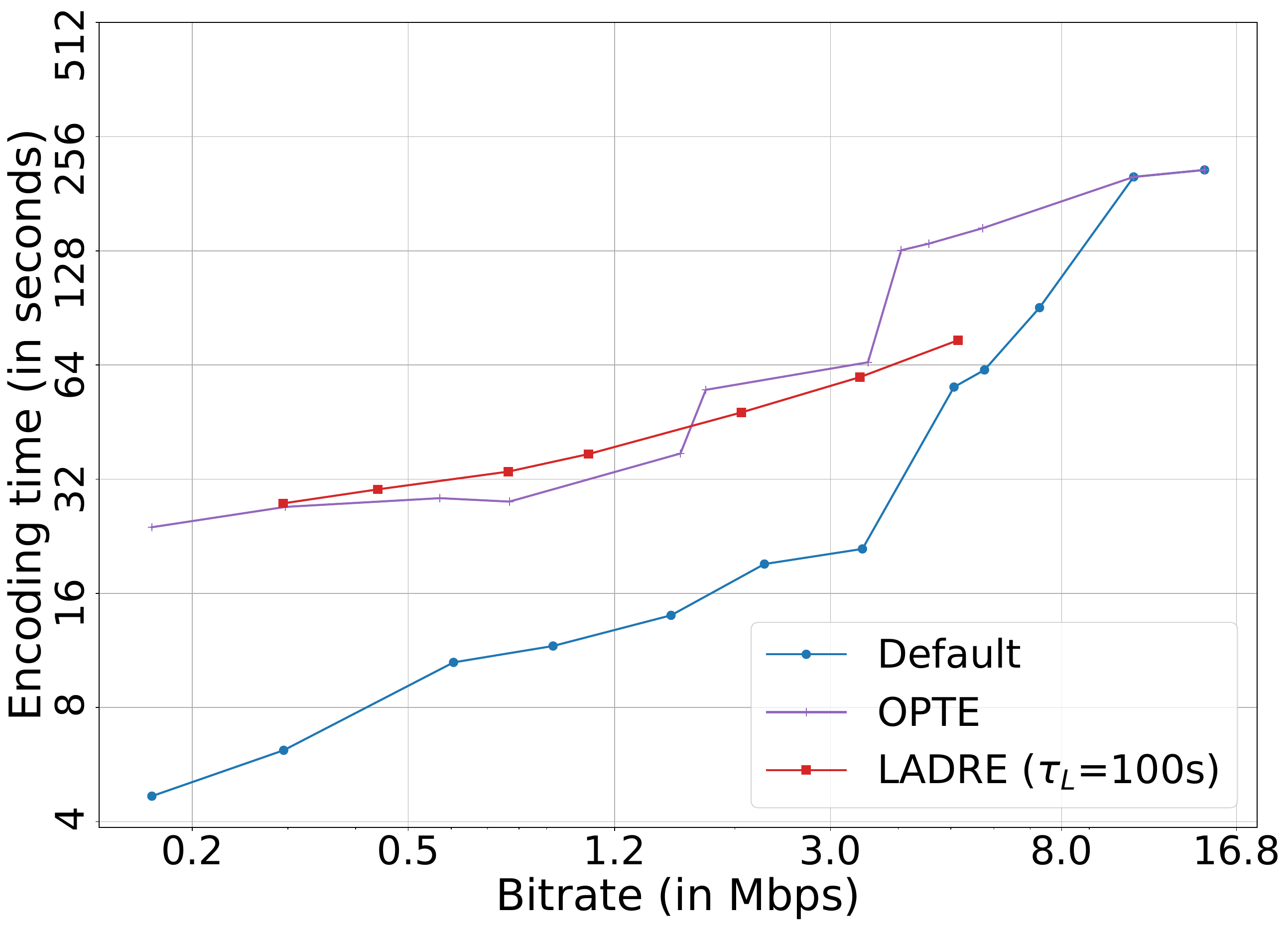}   
    \caption{\textit{0721}}
\end{subfigure}
\caption{RD curves and encoding times of representative video sequences (segments) using default encoding (blue line), \opte (purple line), \scheme (red line).}
\label{fig:rd_res}
\end{figure}

\textit{\textbf{Resolution prediction:}} We analyze the encoding resolution predictions of \scheme. \opte generally yields the highest resolutions for a given target bitrate compared to \textit{Default} and \scheme encodings. The selected encoding resolution for a given target bitrate decreases as $\tau_{\text{L}}$ decreases. If the target latency constraint in \scheme is eliminated, \ie $\tau_{\text{L}}=\infty$, resolutions yielding the highest XPSNR are selected, converging the resolution selection to \opte. Notably, in scenarios where encoding time constraints become more stringent, higher bitrate representations might be omitted in \scheme due to limitations in encoding these representations within the allocated time budget, as observed in Figure~\ref{fig:rd_res}.

\textit{\textbf{Rate-distortion performance:}} Figure~\ref{fig:rd_res} shows the RD curves of the representative video segments in the test dataset. It is observed that the RD curve of \opte is mostly higher than \texttt{default} and \scheme. This means that, for any given target bitrate, \opte maintains a higher level of visual quality as measured by XPSNR. Consequently, viewers can enjoy a visually pleasing experience with reduced artifacts, such as blocking or blurring, at the same bitrate. Notably, some representations are not encoded in \scheme since it was impossible to select a resolution with encoding time within $\tau_{\text{L}}$. Hence, the storage needed for the representations is lowered as $\tau_{\text{L}}$ decreases. It is observed in Table~\ref{tab:res_cons} that the \mbox{BD-PSNR} and \mbox{BD-XPSNR} values decrease as the encoding time constraint is lower. 

\textit{\textbf{Latency and energy consumption performance:}} As shown in Figure~\ref{fig:rd_res}, \opte yields the longest encoding time due to higher encoding resolutions optimized for maximum XPSNR. This significantly increased encoding time may impact real-time or low-latency applications. Encoding typically utilizes the processing units (\eg CPU or GPU) intensively. These processing units operate at a relatively constant power level during encoding. Therefore, the power consumed over time remains reasonably consistent, contributing to the linear relationship between encoding time and energy consumption~\cite{cvfr_ref}. Hence, we assume that the encoding time savings directly translates to the encoding energy consumption reduction. Since we assume that the encodings are carried out concurrently, the total encoding time for each segment ($\tau_{\text{L}}$) is determined to be the highest encoding time yielded among the bitrate ladder representations~\cite{emes_ref}. Table~\ref{tab:res_cons} shows the average encoding time for each segment ($\overline{\tau_{\text{L}}}$) using the considered encoding schemes. It is observed that the encoding times of representations of video segments decrease as $\tau_{\text{L}}$ decreases.

\begin{table}[t]
\caption{Average results of the encoding schemes compared to the \texttt{Default} bitrate ladder encoding.}
\centering
\resizebox{\columnwidth}{!}{
\begin{tabular}{l|c||c|c|c|c|c|c|c}
\specialrule{.12em}{.05em}{.05em}
\specialrule{.12em}{.05em}{.05em}
Method & $\tau_{\text{L}}$ & \makecell{\BDRP} & \BDRV & BD-PSNR & BD-XPSNR &  $\Delta S$ &  $\Delta T \approx \Delta E$ & $\overline{\tau_{\text{L}}}$ \\
& [s] & [\%] & [\%] & [dB] & [dB] & [\%] & [\%] & [s] \\
\specialrule{.12em}{.05em}{.05em}
\specialrule{.12em}{.05em}{.05em}
\opte & - & -28.02\% & -20.50\% & 1.23 & 0.83 & 2.14 & 45.64 & 294.41\\
\hline
\multirow{4}{*}{\scheme} & 50s  & 14.19\%   & 12.76\% & -0.50 & -1.29 & -85.32 & -91.15 & 47.93\\
 &  100s & 6.77\% & 4.49\% & -0.12 & -0.25 & -74.01 & -88.53 & 83.63\\
 &  200s & -10.25\% & -12.03\% & 0.58 & 0.43 & -62.48 & -84.17 & 152.38\\
 &  400s & -21.06\% & -14.51\% & 1.02 & 0.69 & -59.03 & -81.20 & 190.06\\
\specialrule{.12em}{.05em}{.05em}
\specialrule{.12em}{.05em}{.05em}
\end{tabular}
}
\label{tab:res_cons}
\end{table}

\section{Conclusions and future directions}
\label{sec:conclusion}
This paper proposes an encoding latency-aware dynamic resolution per-title encoding scheme (\scheme) for adaptive streaming applications. \scheme includes an optimized resolution prediction, which uses random forest-based models to estimate bitrate-resolution-rate factor triples for a given video segment based on its spatial and temporal characteristics. The experimental results show that, on average, \scheme, with a target encoding time constraint of \SI{200}{\second}, yields bitrate savings of \SI{10.25}{\percent} and \SI{12.03}{\percent} to maintain the same PSNR and XPSNR, respectively, compared to the reference HLS bitrate ladder with a negligible additional latency in streaming. This is accompanied by an average decrease of \SI{84.17}{\percent} in encoding energy consumption.

One promising avenue for future research is exploring advanced machine-learning models to enhance prediction accuracy. Investigating novel features and metrics that better capture the relationship between encoding time and optimal resolutions might also be a promising avenue. Moreover, delving into collaborative frameworks or distributed algorithms for efficient encoding resolution selection across multiple streaming nodes could be another area of exploration.

\balance
\bibliographystyle{IEEEtran}
\bibliography{references.bib}
\balance
\end{document}